\documentclass[aps,reprint,amsmath,
amssymb,eqsecnumaps,prl
 ]{revtex4-2}
\usepackage{graphicx}
\usepackage{dcolumn}
\usepackage{bm}
\usepackage{float}
\usepackage{xcolor}

\newcommand{\p}{\partial}

\newcommand{\om}{\omega}
\newcommand{\vg}{\textsl{g}}
\newcommand{\vD}{\varDelta}

\newcommand{\ta}{\theta}
\newcommand{\sq}[1]{\sqrt{\smash[b]{#1}}}
\newcommand{\al}{\alpha}
\newcommand{\wt}{\widetilde}
\newcommand{\wh}{\widehat}

\newcommand{\cH}{{\cal H}}
\newcommand{\cF}{{\cal F}}

\newcommand{\cW}{{\cal W}}

\newcommand{\be}{\begin{equation}}                                       
\newcommand{\ee}{\end{equation}}
\newcommand{\ba}{\begin{eqnarray}}
\newcommand{\ea}{\end{eqnarray}}
\newcommand{\bref}[1]{(\ref{#1})}

\newcommand{\lab}[1]{\label{#1}}

\newcommand{\tom}{\widetilde{\omega}}
\newcommand{\Omeg}{\textsl{g}}

\newcommand{\kp}{\tfrac{1}{2}\kappa}
\newcommand{\bl}{\begin{align}}
\newcommand{\el}{\end{align}}

\begin{document}

\preprint{APS/123-QED}

\title{Threshold of complexity and Arnold tongues in Kerr ring microresonators}%

\author{D.V. Skryabin}
\email{d.v.skryabin@bath.ac.uk}
\author{Z. Fan}
\author{A. Villois}
\author{D.N. Puzyrev}
\affiliation{Department of Physics, University of Bath, Bath BA2 7AY, UK}%

\date{submitted  04 April, 2020; published January 2021}

\begin{abstract}
We show that the threshold condition for two pump photons to convert into a pair of the sideband ones in Kerr microresonators with high-quality factors breaks the pump laser parameter space into a sequence of narrow in frequency and broad in power  Arnold tongues. Instability tongues become a dominant feature in resonators with the finesse dispersion parameter close to and above one. As pump power is increased, the tongues expand and cross by forming a line of cusps, i.e., the threshold of complexity, where  more sideband pairs become unstable. We elaborate theory for the tongues and threshold of complexity, and report  the synchronisation and frequency-domain symmetry breaking effects inside the tongues.
\end{abstract}

\maketitle

Arnold tongues \cite{arnold1,arnold2} is a well-known phenomenon in parametrically driven  and  
coupled  oscillator systems having cross-disciplinary applications \cite{bio}. The tongues appear as a sequence of the expanding instability and 
synchronisation intervals in the parameter space of the external drive  frequency and amplitude.
The Arnold-tongues concept established itself in  
neuroscience \cite{sin2}, structural \cite{hunt} and nano-mechanics \cite{sin1},  quantum engineering \cite{sin3,sin4,sin5} and in other areas. 
Periodicity embedded into the model
equations is the key property underpinning the formation of Arnold tongues \cite{arnold1,arnold2}.
Nonlinear effects in periodic optical systems 
have traditionally attracted 
considerable  attention \cite{kiv,kr,new}.
In particular, resonators that recirculate light and reinforce 
light-matter interaction are naturally suited to explore the interplay of periodicity, dissipative and nonlinear effects \cite{new,kudl,tur,rev3,rev4,rev5,rev8}.

Arnold tongues emerge through the transformation of  resonances of the underlying linear oscillator  under the influence of dissipative and nonlinear effects. Hence, practical utilisation of this concept in resonators requires  a robust control over the linewidth and mode-density. High-quality-factor microresonators are currently used in the state-of-art frequency conversion, precision spectroscopy, comb generation and single quanta manipulation \cite{rev3,rev4,rev5,rev8,sipe}. 
These devices are few mm to few hundreds of nm long with the tens of GHz to THz resonance separations, i.e., free spectral ranges (FSRs), and finesses $10^3$-$10^6$, that should be compared
to the   finesses $\lesssim 10^2$ in fibre resonators \cite{kudl,tur} and 
other modelocking devices \cite{ursula}.
High finesse is naturally accompanied by the relatively large finesse dispersion, 
which is a measure of the spectral non-equidistance and of the resonance density in the rotating frame, that play a critical role in bringing the microresonator Arnold-tongues to the physical realm.

The Lugiato-Lefever  equation (LLE) \cite{ll}  is a key model in the  microresonator area~\cite{rev3}. Originally proposed in the context of pattern formation in an optical resonator supporting a single longitudinal  and many transverse modes, it has since gained broad interdisciplinary significance \cite{book2}. 
A microresonator implementation of LLE~(mLLE) describes the interaction of many longitudinal modes and therefore should reflect on the aforementioned connection between finesse and dispersion. Exploring an interplay of the two, we have found a broad range of parameters where a transition from the single-mode, i.e., continuous-wave~(cw),  operation to frequency conversion happens via the Arnold-tongue route. The mLLE Arnold-tongues structure  emerges across the pump frequency range from the limit of the infinitesimal pump powers. Therefore a focus of this work is outside the hard excitation regimes, i.e., 
 proximity of the upper state of the bistability loop, giving rise to the mLLE soliton modelocking~\cite{rev3,herr,chembo1,gaeta00,taheri,kart,cole}.

We show that the tongue tips are well separated if the finesse dispersion is the order of one, which implies that the regions of stable single-mode operation are first interleaved with 
the instability intervals where only one pair of sidebands with the mode numbers $\pm\mu$ kicks off the frequency conversion process. As the pump is increased the tongues expand, at some point they start overlapping and form the complexity threshold. Above this threshold, two and more pairs of modes with different $|\mu|$ are becoming unstable simultaneously. We elaborate a theory of the threshold of complexity and report nonlinear effects in its proximity. 
The term - {\em threshold of complexity} is inspired by the cellular automata related concepts and terminology~\cite{compl}. 

We now introduce a model and explain how it  maps onto physical devices,
while more cross-area context is provided through the text and before summary.
Amplitude $\psi$ of the electric field in a ring microresonator can be expressed as  
a superposition of angular harmonics: $\psi=\sum_\mu\psi_\mu(t) e^{i\mu\vartheta}$.
Here  $\mu$ and $\psi_\mu$ are the mode numbers and  amplitudes. 
$\vartheta\in[0,2\pi)$ is an angle varying along the ring. 
Resonant frequencies, $\omega_\mu$, are counted from  $\om_0$ and approximated as 
\begin{equation}
	\om_\mu=\om_0 +D_1\mu+\tfrac{1}{2}D_2\mu^2,~ \mu=0,\pm 1,\pm 2,\dots .
\lab{freq}
\end{equation}
$D_1/2\pi$ is the resonator FSR, and  
$D_2$ is the second order dispersion, characterising how FSR is changing with $\mu$.
For example, $D_1/2\pi= 15$GHz, $|D_2|/2\pi=1$kHz are good estimates for CaF$_2$  resonators  for $\omega_0/2\pi$  
being close to $200$THz pump, $\om_p$, \cite{hq1}. If $\kappa/2\pi$ is the linewidth, then  
the resonator finesse is 
\begin{equation}
\cF_\mu=\pm\frac{\omega_{\mu\pm 1}-\omega_{\mu}}{\kappa}=
\cF+\al_\mu,~ 
\cF=\frac{D_1}{\kappa},
\lab{q}
\end{equation}
where we take $+1$ for $\mu>0$ and $-1$ for $\mu<0$. Here, $\cF$ is
the dispersion free finesse, and  $\al_\mu=\cF_\mu-\cF$ is the residual finesse,
\be
\al_\mu=\big(\mu\pm\tfrac{1}{2}\big)\cF_d\approx\mu\cF_d,~ \cF_d=\frac{D_2}{\kappa}.
\lab{alpha}
\ee
$\cF_d$ is the finesse dispersion, which is a 
key parameter in what follows. It can be either positive (anomalous dispersion, $D_2>0$)
or negative (normal dispersion, $D_2<0$). 
$Q=\om_0/\kappa$ is the resonator quality factor. 
$Q=3\cdot 10^9$, which is a conservative number for CaF$_2$ resonators, gives 
$\kappa/2\pi= 67$kHz, $\cF=22\cdot 10^4$, and
$|\cF_d|\lesssim 1.5\cdot 10^{-2}$ for these samples  \cite{hq1,hq2}. For the high  $Q=10^{11}$ samples \cite{hq3}, $\kappa/2\pi= 2$kHz, $\cF=750\cdot 10^4$, and $|\cF_d|\lesssim 0.5$.
 
The Kerr mLLE  is
\cite{herr,chembo1,dvs} 
\be
i\p_t\psi=\delta_0\psi-\tfrac{1}{2}D_2\p_\ta^2\psi-i\tfrac{1}{2}\kappa
\left(\psi-\cH \right) 
-\gamma |\psi|^2\psi.
\lab{ll}\ee
Here, $\cH^2=\frac{\eta }{\pi}\cF\cW$ is the intracavity pump power, 
where $\cW$ is the laser power.
$\eta=\kappa_c/\kappa<1$  is the pump coupling efficiency and $\kappa_c$ is the  coupling rate. $\gamma/2\pi= 10$kHz/W is the nonlinear parameter  \cite{dvs}. 
$\delta_0=\om_0-\om_p$ is detuning of the pump laser frequency, $\om_p$,
from  $\om_0$.  

Transformation to the rotating reference frame, $\ta=\vartheta-D_1t$, 
replaces the  $\om_\mu$-spectrum with  the residual, i.e.,   mLLE, spectrum,
\be
\tfrac{1}{\kappa}{\vD_\mu}=\tfrac{1}{\kappa}{\delta_0}+\tfrac{1}{2}\mu^2\cF_d.
\lab{res}
\ee 
Respectively, the 
finesse $\cF_\mu$ is replaced with the residual finesse, $\al_\mu$, Eq. \bref{alpha}. The latter is also expressed as $\al_\mu=\pm(\vD_{\mu\pm 1}-\vD_\mu)/\kappa$ and can be interpreted as the inverse mode density. Fig.  1(c) illustrates how the density of states in the residual spectrum is reduced with $\cF_d$ increasing, and also shows that the linear resonances, $\vD_\mu=0$, are located at $\delta_0<0$ for $\cF_d>0$, and at $\delta_0>0$ for $\cF_d<0$. 
Residual spectrum is non-equidistant, and hence,  even for very small $\cF_d$'s there always be a sufficiently large $|\mu|$ making its resonances well separated.  Thus, $|\al_\mu|\ll 1$   and  $|\al_\mu|\gg 1$ correspond to the quasi-continuous and sparse 
limits of the residual spectrum, respectively. Note, that the 
finesse itself is maintained arbitrarily large in any case, $\cF_\mu\gg |\al_\mu|$. 

We now define the single-mode  cw-solution of Eq. \bref{ll}  as
$\psi=\sq{\vg/\gamma}~e^{i\phi_0}$, where $\vg>0$ solves
\be
\gamma\cH^2=\vg+4\vg\left(\delta_0-\vg\right)^2/\kappa^2,
\lab{cw}
\ee
and recapture key steps in its stability analysis \cite{ll}.
Perturbing the cw with a pair of sideband modes, 
$\psi=\sq{\vg/\gamma}~e^{i\phi_0}
+\psi_{\mu}e^{i\mu\ta}+\psi^*_{-\mu}e^{-i\mu\ta}$,
we find $i\p_t\vec q_\mu=\wh V_\mu\vec q_\mu$, 
where $\vec q_\mu=(\psi_{\mu},\psi_{-\mu})^T$, and
\begin{equation}
\wh V_\mu=\begin{bmatrix}
\vD_\mu-2\vg-i\kp & -\vg e^{i2\phi_0}  \\
\vg  e^{-i2\phi_0} & -\vD_\mu+2\vg-i\kp 
\end{bmatrix}.
\lab{mat}
\end{equation}
Thus, $\vg$ is simultaneously responsible for the nonlinear shifts of the resonances and  the anti-Hermitian coupling between the $\pm\mu$ sidebands.
Setting $\vec q_\mu\sim e^{\lambda_\mu t}$ gives \cite{ll}
\begin{eqnarray}
&& \lambda_\mu\big(\lambda_\mu+\kappa\big)=
3\big(\Omeg^{(1)}_{\mu}-\Omeg\big)\big(\Omeg-\Omeg^{(2)}_{\mu}\big),\lab{lam}\\
&&\Omeg^{(1),(2)}_{\mu}=\tfrac{2}{3}\vD_\mu\pm\tfrac{1}{3}
\sqrt{\smash[b]{\vD_\mu^2-\tfrac{3}{4}\kappa^2}}.
\lab{thr}
\end{eqnarray}

Degenerate four-wave-mixing (FWM) process described by $\wh V_\mu$ corresponds to the following photon energy conservation: 
$2\hbar\om_p=\hbar(\om_p+\mu D_1+\text{Im}\lambda_\mu)+\hbar(\om_p-\mu D_1-\text{Im}\lambda_\mu)$. CW stability is neutral at $\text{Re}\lambda_\mu=\lambda_\mu=0$, 
\be
\vg=\vg^{(1)}_\mu, \vg=\vg^{(2)}_\mu. 
\lab{t1}
\ee
Hence, FWM sidebands are exponentially amplified for  $\vg^{(2)}_\mu<\Omeg<\vg^{(1)}_\mu$. Lower boundary of the cw stability in $(\delta_0,\cH)$-plane, i.e., the FWM-threshold, is made by the minima of  
$\vg^{(2)}_\mu$ in $\delta_0$, which are found at  $\vD_\mu=\kappa$, $\vg_\mu^{(2)}=\kappa/2$ for every $\mu$.
Explicitly, the corresponding intracavity pump power, $\cH_{\mu F}^2$, and 
detuning, $\delta_0^{(\mu F)}=\om_0-\om_p^{(\mu F)}$, along the  
FWM threshold is
\begin{eqnarray}
&&\gamma\cH_{\mu F}^2=\tfrac{1}{2}\kappa\big[1+(1-\mu^2\cF_d)^2
\big],
\lab{lp}\\
&&\tfrac{1}{\kappa}\delta_0^{(\mu F)}=1-\tfrac{1}{2}\mu^2\cF_d.
\lab{lf}
\end{eqnarray}
Eliminating $\mu^2$ one finds 
\begin{equation}
\gamma \cH_{\mu F}^2=\tfrac{1}{2}\kappa
\big[1+\big(1-\tfrac{2}{\kappa}\delta_0^{(\mu F)}
\big)^2\big]
\lab{e011}
\end{equation}
For $\kappa=2$, Eqs. \bref{lf} and \bref{e011} match equations for $a(n_c)$ and  $E_{Ic}$ from Ref.~\cite{ll}. Here and below,  the sub- and super-scripts $F,C$  stand for the 'FWM' and 'Complexity' thresholds, respectively.

\begin{figure*}[t]
	\centering{
		\includegraphics[width=0.99\textwidth]{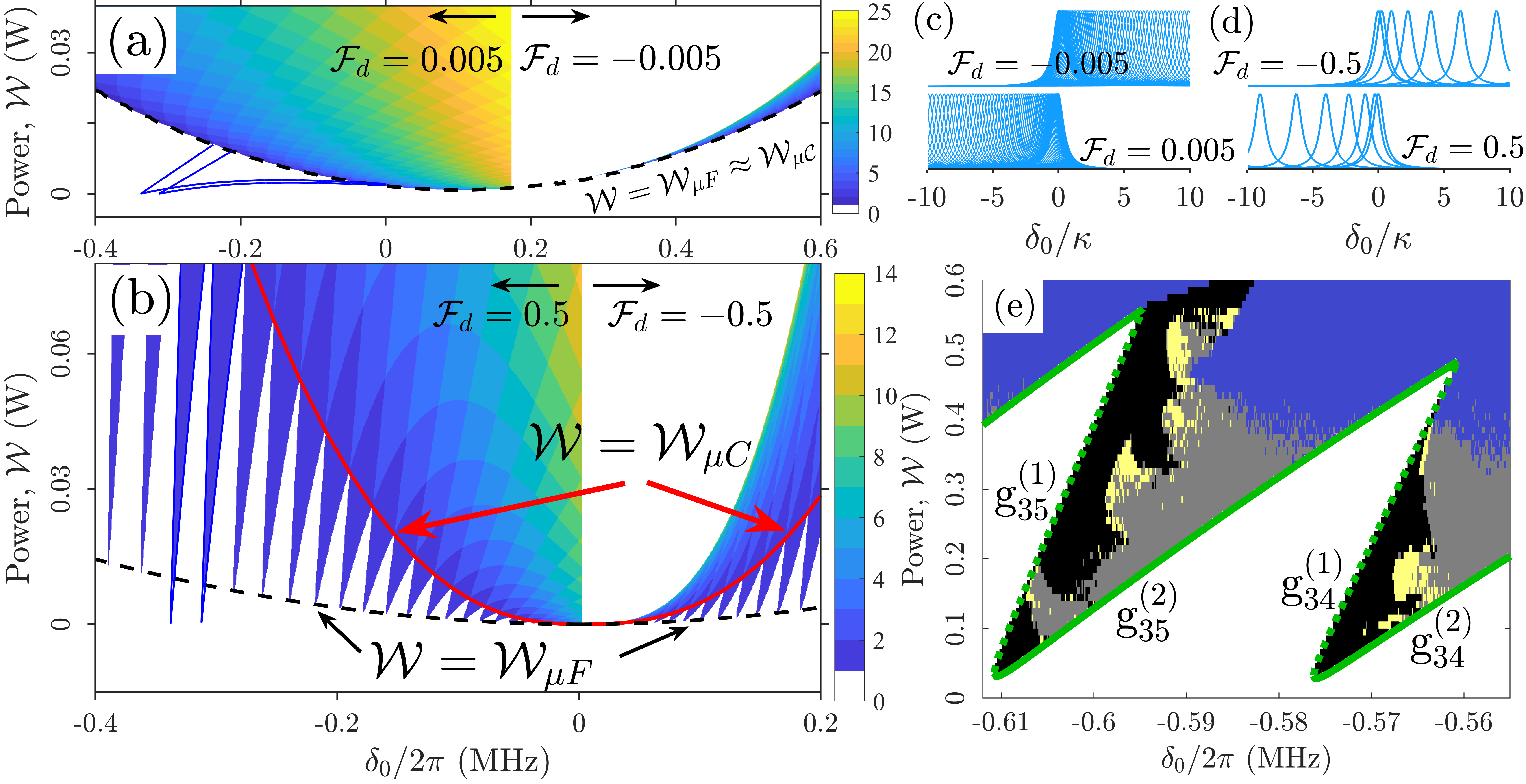}
	}		
	\caption{ 
		{\bf (a,b)}~Arnold tongues  and interplay of the FWM, $\cW=\cW_{\mu F}$, and complexity,  $\cW=\cW_{\mu C}$, power thresholds for the quasi-continuous
		($|\cF_d|=0.005$, $Q=10^{9}$)  and sparse 
		($|\cF_d|=0.5$, $Q=10^{11}$) residual spectra; $|D_2|/2\pi=1$kHz, $\eta=0.5$, residual finesse $\cF_d=D_2/\kappa=D_2Q/\om_0$.  
		The color scheme shows the number of the simultaneously 
		unstable $\pm\mu$ sideband pairs. Lines dropping below $\cW=\cW_{\mu F}$ and reaching $\cW=0$ show a pair of the synchronisation tongues. 
		{\bf (c,d)}~Resonances in the residual spectrum for small (c) and large (d) $|\cF_d|$. Plots show Lorentzian lines, $1/(1+4\vD_\mu^2/\kappa^2)$, for $\mu$ starting from $0$. 
		{\bf (e)}~Distribution of the dynamical regimes across the $|\mu|=35$ instability tongue and its 
		neighbourhood for $\cF_d=0.5$. Black color corresponds to the repetition rate locking to $|\mu|D_1$, see Fig.~\ref{f2}(a). 
		Yellow marks the 'unlocked' regimes with the $|d_\mu|=\text{const}$ symmetry breaking, see Fig.~\ref{f2}(b). Grey is the weak chaos with the symmetry breaking, see Fig.~\ref{f2}(c). Blue is the multimode complexity with dense continuous spectra, see Fig.~\ref{f2}(d). The intra-resonator cw power at the tongue minima $\vg_{35}^{(2)}/\gamma=\kappa/2\gamma=0.1$W (FWM threshold), and 
		$\vg_{35}^{(1)}/\gamma=\vg_{36}^{(2)}/\gamma\simeq \alpha_{35}\kappa/2\gamma=1.75$W at the  threshold of complexity.
	}
	\label{f1}
\end{figure*}

Eqs. \bref{lp}, \bref{lf} represent a discrete set of values, while the laser power and frequency can be tuned continuously. 
Hence, the instability threshold in $(\cH,\delta_0)$  also exists between 
the discrete points with the coordinates specified by $\delta_0^{(\mu F)}$ and 
$\cH_{\mu F}^2$.  
To  demonstrate if and when their separation is important,  
we substitute $\vg=\vg_\mu^{(1)}$ and $\vg=\vg_\mu^{(2)}$   
directly to Eq. \bref{cw} and compute $\cH^2$ vs $\delta_0$ for all $\mu$.  This procedure returns, in general, 
an infinite number, one for each $|\mu|$, of threshold lines that are shown in Figs.~\ref{f1}(a),(b)
for $|\cF_d|=0.005$ and  $0.5$ ($Q=10^9$ and $10^{11}$). 
For relatively small $|\cF_d|$ and quasi-continuous residual spectra, Fig. \ref{f1}(c), 
the individual instability lines in Fig. \ref{f1}(a) overlap 
tightly and form a visibly single threshold \cite{chembo1} reproduced by Eq. \bref{e011} with $\cH_{\mu F}$ vs $\delta_0^{(\mu F)}$ considered as a continuous function.

As $|\cF_d|$ starts approaching $1$ and if it goes above it, then resonances in the residual spectrum, Fig. \ref{f1}(d), and the respective thresholds for the neighbouring $|\mu|$  separate, so that the pump frequency and power range with FWM gain reshapes  profoundly and forms instability tongues, Fig. \ref{f1}(b). Now, Eq. \bref{e011} describes only the lowest power limit at which instability  becomes possible, with the large areas of stability present above it, see dashed black line in Fig.~\ref{f1}(b).   We note, that $\om_p$, in Figs. \ref{f1}(a,b), is scanned across the narrow interval of  $10^{-4}D_1$ around $\om_0$ so that $\om_{\mu\ne 0}$ are not approached even remotely.

The instability tongues, when they are formed, provide selective excitation conditions for the sideband pairs with a given $|\mu|$.  The tongue tips have been found along the quasi-linear tails of the $\vg$~vs~$\delta_0$  solution of Eq.~\bref{cw} for $\delta_0<\sqrt{3}\kappa/2=\delta_b$  if $\cF_d>0$, 
and  for  $\delta_0>\delta_b$ if $\cF_d<0$. The tips are thus directly connected to the $\vD_\mu=0$ resonances in the residual spectrum of the linear, $\vg\to 0$, resonator, see Fig.~\ref{f1}, Eq.~\bref{res}, and more details below.  The limiting value of $\delta_0=\delta_b$, where the tongue structure diminishes, is found from $\vg_0^{(1)}=\vg_0^{(2)}$, which is the condition for the onset of bistability and   the soliton regime ($\cF_d>0$, $\delta_0>\delta_b$) \cite{rev3}.  Tongues have also been found by us in Si$_3$N$_4$ 
resonators~\cite{ff1} with $Q\sim 10^6-10^7$~\cite{hq4,hq5},
and $D_2\sim 10^4$kHz. The $\vg=\vg_\mu^{(1)}$ (green squares in Fig.~\ref{f1}(e)) and $\vg=\vg_\mu^{(2)}$ (green full lines) conditions shape  
the edges of the tongues, while their tips  always belong to $\vg=\vg_\mu^{(2)}$. 

Importantly, there is also a well-defined line that limits the area above 
the tongue tips and below their intersections, see red line in Fig.~\ref{f1}(b). We call this boundary - the threshold of complexity. 
Reaching this threshold signals two events. First is that the inter-tongue stability intervals 
cease to exist.  Second is that the resonator is brought into the regime
where $\pm\mu$ and $\pm(\mu+1)$ side-bands can become simultaneously unstable. 
As the pump is increased further a sequence of thresholds, where more modes 
become unstable,  can be seen in Figs.~\ref{f1}(a),(b). For smaller 
$|\cF_d|$ the residual spectrum is dense and all the thresholds tend to merge, 
while,  for larger $|\cF_d|$, their separation is pronounced. 

Thus, the complexity threshold consists of the points where  bifurcation lines 
corresponding to the excitations of the $\pm\mu$-sidebands 
intersect with the ones for  $\pm(\mu+1)$. Since $\vD_\mu$ is a function of $\mu^2$,
assuming $\mu>0$ does not restrict the generality.
Now, the intersection points, for $\cF_d>0$, can be found applying 
	\be    
\vg=\vg_{\mu}^{(2)}=\vg_{\mu+ 1}^{(2)},	
~\text{and}~
\vg=\vg_{\mu}^{(1)}=\vg_{\mu+ 1}^{(2)},
	\lab{co2}
\ee
cf., Eq.~\bref{t1}. If $\cF_d<0$, then the second condition becomes
$\vg=\vg_{\mu}^{(2)}=\vg_{\mu+ 1}^{(1)}$.  Equations~\bref{co2} is the  mathematical
quintessence of the complexity threshold.
Each of Eqs.~\bref{co2} is a double, i.e., codimension-2,  
condition which marks a sequence of  the
instability points along the threshold line, see Fig.~\ref{f1}(b). 
Equations~\bref{co2} are the cusp conditions and hence are non-differentiable in $\delta_0$, unlike  $\vg=\vg_\mu^{(2)}$, Eqs.~\bref{t1}. 
For $|\al_\mu|\ll 1$, $|\cF_d|\ll 1$, the cusps become very shallow and the tongue structures disappear, while $\cF$ is still $\gg 1$.

\begin{figure*}[t]
	\includegraphics[width=0.95\textwidth]{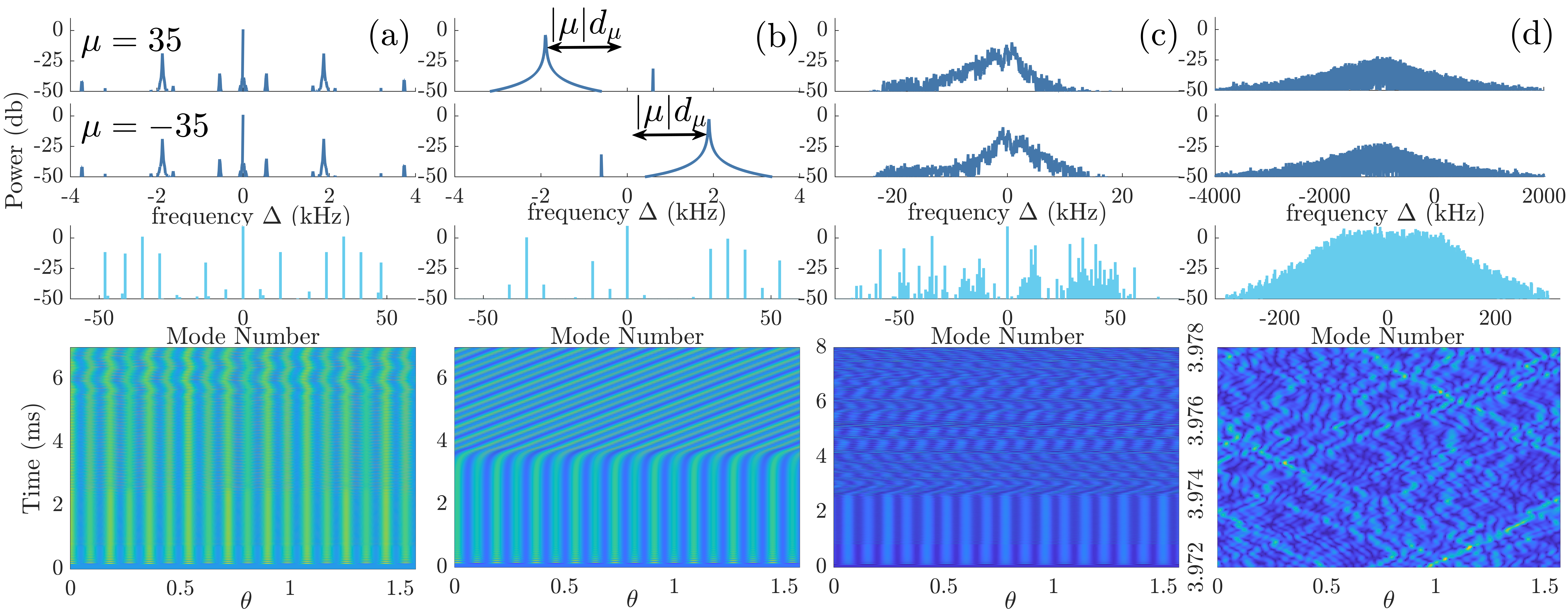}
	\caption{{\bf (a)}~Repetition rate locking ($d_\mu=0$, black area in Fig.~1(e)); {\bf (b)} Frequency-domain symmetry breaking ($d_\mu=\text{const}$, yellow area in Fig.~1(e)); {\bf (c)}~Few-mode chaos (chaotic variations of $d_\mu$, grey area in Fig.~1(e))
		{\bf (d)}~Multi-mode complexity (blue area in Fig.~1(e)). Top two rows show the $\mu=\pm 35$ sideband RF spectra $|S_{\mu}(\Delta)|^2$~\cite{beat}.   3rd row shows the 
		mode powers at $t=16$ms. 4th row is the space-time dynamics. 
	}
	\label{f2}
\end{figure*}

Resolving Eqs. \bref{co2}  
gives 
\be
\vD_\mu=\kappa\big(-\tfrac{1}{2}\al_\mu+\sqrt{\smash[b]{1+\al_\mu^2}}
~\big).
\lab{cusp1}
\ee
The respective pump detunings are found applying Eqs.~\bref{res},~\bref{cusp1},
\be
\tfrac{1}{\kappa}\delta_0^{(\mu C)}=
\big(-\tfrac{1}{2}\al_\mu+\sqrt{\smash[b]{1+\al_\mu^2}}
~\big)-\tfrac{1}{2}\mu^2\cF_d,
\lab{Uf}
\ee
cf., Eq. \bref{lf}.

If, e.g.,  $\al_\mu\gg 1$ ($\mu\gg 1$, $\cF_d>0$), then 
$-\tfrac{1}{2}\al_\mu+\sqrt{\smash[b]{1+\al_\mu^2}}
=\tfrac{1}{2}\al_\mu+{\cal O}
\big(\al_\mu^{-1}\big)$.
Hence, Eq.~\bref{thr} gives $\vg=\vg_\mu^{(1)}\simeq \tfrac{1}{2}\kappa \al_\mu $ along the threshold of complexity.
Simultaneously,  the relatively large detunings, $|\delta_0|\gg \vg$, imply dispersive quasi-linear resonator response. E.g., for $\kappa/2\pi=2$kHz ($Q=10^{11}$) and $\mu=35$, we have $\delta_0/2\pi\simeq 0.5$MHz and $\vg=\vg_{35}^{(1)}\simeq 2\pi\times 17$kHz at the complexity threshold.
Therefore, Eq.~\bref{cw} can be approximated by $\gamma\cH^2\simeq 4\Omeg\delta_0^2/\kappa^2$ and, hence, the laser power  is 
\be
\cW_{\mu C}\simeq \frac{\pi}{\eta\cF}\frac{\kappa\mu^2}{2\gamma} |\al_\mu|^3=
\frac{\pi D_2^2}{2\gamma D_1}~\frac{|\mu D_2|}{\eta\kappa}\mu^4,~|\al_\mu|\gg 1. 
\lab{Uw}  
\ee

$\vD_\mu=\kappa$ and $\vg=\vg_\mu^{(2)}=\kappa/2$ at the FWM threshold, see Eq. \bref{e011},
 and hence $\gamma\cH^2\simeq 4\Omeg\delta_0^2/\kappa^2$ now gives 
 $\cW_{\mu F}=\cW_{\mu C}/|\al_\mu|=
\kappa\cW_{\mu C}/|\mu D_2|$.
The prefactor 
that makes the difference between the powers at the FWM and complexity thresholds and hence measuring 
the relative depth of the instability tongues  is exactly 
the residual finesse, $\al_\mu$. 

The proximity of each tongue provides a parameter range where the mLLE dynamics 
is dominated by the competition between the two normal modes 
of $\wh V_\mu$, $\wh V_\mu\vec q_\mu^{(\pm)}=i\lambda_\mu^{(\pm)}\vec q_\mu^{(\pm)}$. Rearranging 
Eq.~\bref{lam} as
$(\lambda_\mu+\tfrac{1}{2}\kappa)^2=(\vD_\mu-\vg)(3\vg-\vD_\mu)$, we find 
$\text{Im}\lambda_\mu^{(\pm)}=\pm~\text{Im}
\sq{3(\vD_\mu-\vg)(\vg-\tfrac{1}{3}\vD_\mu)}$, and hence
$\text{Im}\lambda_\mu^{(\pm)}=0$ is satisfied for
$\tfrac{1}{3}\vD_\mu<\vg<\vD_\mu$. This is the normal mode  synchronisation condition implying that
 the repetition rate of the emerging quasi-harmonic in $\ta$, i.e., roll, patterns locks to  $|\mu| D_1$. 
While $D_1$  is a property of the linear resonator and is independent of the pump parameters, $\om_p$ and $\vg$, the normal-mode synchronisation implies that the nonlinear, $\vg\ne 0$,  repetition rate locks to an integer-multiple of $D_1$. The instability and  synchronisation tongues  merge together with $|\cF_d|$ increasing, and $\vg_\mu^{(1)}\to\vD_\mu$,  $\vg_\mu^{(2)}\to\tfrac{1}{3}\vD_\mu$, cf.,  Figs.~\ref{f1}(a) and (b) where two 
synchronisation tongues are contrasted against the instability ones.

Tongues are classified by their synchronisation order
$n~:~m$, where $n$ is the number of nonlinear pulses coming through the system per  $m$ periods of the linear oscillator~\cite{bio}. In our case, the linear round trip time is $2\pi/D_1$ and periods of the synchronised waveforms are $2\pi/|\mu|D_1$, thus the corresponding orders are $|\mu|~:~1$.
As for the classic Arnold tongues~\cite{bio}, the synchronization intervals start for infinitesimal pumps at the resonance points, $\vD_\mu=0$, in the linear, $\vg\to 0$, spectrum. The upper and middle branches of the $\vg$ vs $\delta_0$ curve inside the bistability interval  (unlike its quasi-linear tails) do not withstand the $\vg\to 0$ limit, and therefore their stability maps do not reveal the tongues.

Though the linear theory predicts $|\mu|~:~1$ synchronisation across the whole tongues area, the nonlinear processes make other regimes possible.
The time-average mode frequencies are  $\wt \om_{\mu}=\omega_p+D_1\mu+\langle\p_t\phi_\mu\rangle$, where $\phi_\mu(t)=\text{arg}\psi_\mu$~\cite{beat}.
Beating  $\psi_\mu$ against $\psi_{-\mu}$ provides a measure
of the average nonlinear repetition rate
$\wt  D_{1\mu}=\frac{1}{2}(\tom_{\mu }-\tom_{-\mu })$. The difference between the two rates, $\wt  D_{1\mu}-|\mu|D_1
=\frac{1}{2}(\langle\p_t\phi_{\mu }\rangle-\langle\p_t\phi_{-\mu }\rangle)\equiv|\mu|d_{\mu}$, is a function of $\om_p$, and $\vg$. Hence, $d_\mu\ne 0$ implies de-synchronisation, i.e., 
'unlocking' of the repetition rate from $|\mu|D_1$.

Numerical simulations of Eq.~\bref{ll} conducted for $20000$ parameter points
across the tongues have revealed that the locking, $d_\mu=0$, range 
starts from the tongue tips and  spreads along their $\vg=\vg_\mu^{(1)}$ edges  upwards, see black areas in Fig.~\ref{f1}(e), and Fig.~\ref{f2}(a). 
Another expressed feature of the dynamics inside the tongues is the violation of the repetition rate locking regimes producing $d_\mu\ne 0$, see yellow and grey areas in Fig.~\ref{f1}(e). 
In the coordinate space, the $d_\mu\ne 0$ regimes correspond to a pair of the coexisting rolls rotating with the unlocked  rates,  $|\mu|(D_1\pm  d_{\mu})$. Two rolls can either coexist independently (yellow areas), see one of them in Fig.~\ref{f2}(b), or  generate chaotic switching dynamics (grey areas) associated with the weak modes emerging around the dominant ones, see Fig.~\ref{f2}(c).
Here we deal with the frequency-domain symmetry breaking~\cite{skr}, when 
symmetry is broken not only for the sideband powers, see mode-number spectra
in  Figs.~\ref{f2}(b), but also for their spectral content, see and compare the radio-frequency (RF) spectra of the dominant modes in  Figs.~\ref{f2}(a) and \ref{f2}(b).
As the threshold of complexity
is approached and crossed, the few-mode dynamics is replaced by the transition to the 
developed multimode chaos, when every consecutive mode in the spectrum is excited forming  very dense continua, see Fig.~\ref{f2}(d).

Before concluding, we make further  contextual connections. Refs.~\cite{gen1,gen2} looked at the similar to our
	 problem of the Arnold tongue overlaps  in dissipative maps and  biological systems. 
Refs.~\cite{gaeta0,gaeta} measured synchronisation and tongues for the soliton sequences flowing across a fibre linking two micro-rings. 
Tongue formation studied by us should be compared  with the multiple instability domains containing continua
 of modes reported in the zero finesse diffractive feedback
systems~\cite{firth}, and in the low-finesse resonators~\cite{tur,kudl,kar}.
Refs.~\cite{ll2,tp3,oppo,chembo2,optica} reported stationary and breathing rolls with no symmetry breaking in mLLE and alike systems.
Refs.~\cite{skr,pas,kip} studied symmetry breaking of the counter-propagating single-mode and soliton regimes in bi-directional microresonators.

In summary: We elaborated theory for the threshold of complexity  and  nonlinear effects within Arnold tongues in Kerr microresonators. 
Our results hold potential for applications in frequency conversion and RF-photonics areas relying on a range of existing and emerging high-Q resonators.

\begin{acknowledgments}
This work was supported by the EU  Horizon 2020 
Framework Programme (812818, MICROCOMB). 
We are deeply indebted to  
W.J. Firth and L.A. Lugiato for invaluable comments and interest  
in our work.
\end{acknowledgments}


\begin{thebibliography}{9}
\bibitem{beat}
$\langle\p_t\phi_{\mu }\rangle=\int\Delta|S_\mu|^2 d\Delta/\int |S_\mu|^2 d\Delta$,
where $S_\mu(\Delta)=\tau^{-1}\int_{\tau_1}^{\tau_1+\tau} \psi_\mu(t) e^{i\Delta t}dt$. 
The values of $d_\mu$ used  to  separate the yellow and grey areas
in Fig. \ref{f1}(e) were  also averaged over $\tau_1$.
\end{thebibliography}
\end{document}